\documentclass[journal]{IEEEtran}

\ifCLASSINFOpdf

\else

\fi

\usepackage{nomencl}
\usepackage{amsmath}
\usepackage{amssymb}
\usepackage{color}
\usepackage[cmyk]{xcolor}
\usepackage{graphicx}
\usepackage{makecell}
\usepackage{multirow}
\usepackage{diagbox}
\usepackage{cite}
\usepackage{stfloats}
\usepackage[normalem]{ulem} 
\graphicspath{ {./images/} }
\makenomenclature

\usepackage{etoolbox}
\renewcommand\nomgroup[1]{%
  \item[\bfseries
  \ifstrequal{#1}{A}{Acronyms}{%
  \ifstrequal{#1}{I}{Indices and Sets}{%
  \ifstrequal{#1}{P}{Parameters}{%
  \ifstrequal{#1}{V}{Variables}{%
  }}}}%
]}

\begin{document}
\title{A Robust Planning Model for Offshore Microgrid\\Considering Tidal Power and Desalination}
\author{Zhimeng~Wang,
        Ang~Xuan,
        Xinwei~Shen,
        Yunfei~Du,
        and~Hongbin~Sun
\thanks{This work was supported by National Natural Science Foundation of China (No.52007123). (Corresponding: Xinwei Shen, xwshen@tsinghua.edu.cn)}
\thanks{Z. Wang, A. Xuan, X. Shen, and Y. Du are with Tsinghua Shenzhen International Graduate School (SIGS), Tsinghua University, Shenzhen, 518055, China. H. Sun is with Taiyuan University of Technology, Taiyuan, 030002, China and also with the Department of Electrical Engineering, Tsinghua University, Beijing, 100084, China.}
}

\maketitle

\begin{abstract}
Increasing attention has been paid to resources on islands, thus microgrids on islands need to be invested. 
Different from onshore microgrids, offshore microgrids (OM) are usually abundant in ocean renewable energy (ORE), such as offshore wind, tidal power generation (TPG), etc. 
Moreover, some special loads such as seawater desalination unit (SDU) should be included. 
In this sense, this paper proposes a planning method for OM to minimize the investment cost while the ORE's fluctuation could be accommodated with robustness. 
First, a deterministic planning model (DPM) is formulated for the OM with TPG and SDU. 
A robust planning model (RPM) is then developed considering the uncertainties from both TPG and load demand. 
The Column-and-constraint generation (C\&CG) algorithm is then employed to solve the RPM, producing planning results for the OM that is robust against the worst scenario. 
Results of the case studies show that the investment and operation decisions of the proposed model are robust, and TPG shows good complementarity with the other RESs. 
\end{abstract}


\nomenclature[A,01]{\(\mathrm{OM}\)}{Offshore Microgrid. }
\nomenclature[A,02]{\(\mathrm{DER}\)}{Distributed Energy Resource. }
\nomenclature[A,03]{\(\mathrm{RES}\)}{Renewable Energy Source. }
\nomenclature[A,04]{\(\mathrm{DU}\)}{Dispatchable Unit. }
\nomenclature[A,05]{\(\mathrm{NDU}\)}{Nondispatchable Unit. }
\nomenclature[A,06]{\(\mathrm{ESS}\)}{Energy Storage System.}
\nomenclature[A,07]{\(\mathrm{RO}\)}{Robust Optimization. }
\nomenclature[A,08]{\(\mathrm{SO}\)}{Stochastic Optimization. }
\nomenclature[A,09]{\(\mathrm{C\&CG}\)}{Column-and-Constraint Generation. }
\nomenclature[A,10]{\(\mathrm{DPM}\)}{Deterministic Planning Model. }
\nomenclature[A,11]{\(\mathrm{RPM}\)}{Robust Planning Model. }
\nomenclature[A,12]{\(\mathrm{ORE}\)}{Ocean Renewable Energy. }
\nomenclature[A,13]{\(\mathrm{SDU}\)}{Seawater Desalination Unit. }
\nomenclature[A,14]{\(\mathrm{TPG}\)}{Tidal Power Generation. }

\nomenclature[I,01]{\(i,\Omega_{DU}\)}{Index and Set for dispatchable units. }
\nomenclature[I,02]{\(j,\Omega_{NDU}\)}{Index and Set for traditional nondispatchable units. }
\nomenclature[I,03]{\(k,\Omega_{TPG}\)}{Index and Set for tidal generation units. }
\nomenclature[I,04]{\(l,\Omega_{ESS}\)}{Index and Set for energy storage systems. }
\nomenclature[I,05]{\(h,H\)}{Index and Set for hour. }
\nomenclature[I,06]{\(d,D\)}{Index and Set for day. }
\nomenclature[I,07]{\(y,Y\)}{Index and Set for year. }
\nomenclature[I,08]{\(\mathcal{U}\)}{Uncertainty Set. }

\nomenclature[P,01]{\(cc\)}{Annualized investment cost of generating units, \$ per MW. }
\nomenclature[P,02]{\(cp\)}{Annualized investment cost of storage (power), \$ per MW. }
\nomenclature[P,03]{\(ce\)}{Annualized investment cost of storage (energy), \$ per MW. }
\nomenclature[P,04]{\(RP_i,RC_i\)}{Rated power and rated capacity of the $i$th generation device, MW. }
\nomenclature[P,05]{\(c_i\)}{Levelized operation cost of the $i$th generation device, \$ per MWh. }
\nomenclature[P,06]{\(dr\)}{Discount rate, dimensionless. }
\nomenclature[P,07]{\(\kappa_y\)}{Coefficient of present-worth value of the $y$th year, dimensionless. }
\nomenclature[P,08]{\(\nu\)}{Punishment for load shedding, dimensionless. }
\nomenclature[P,09]{\(L^{\max}\)}{Maximal predicted load demand, MW. }
\nomenclature[P,10]{\(\tilde{L}_{h,d,y}\)}{Predicted electricity load demand at hour $h$, day $d$, year $y$, MW. }
\nomenclature[P,11]{\(\overline{L},\underline{L}\)}{Upper and lower deviation of electricity load demand, respectively, MW. }
\nomenclature[P,12]{\(RC_F\)}{Rated capacity of the seawater desalination unit, t. }
\nomenclature[P,13]{\(cc_F\)}{Annualized investment cost of desalination unit, \$. }
\nomenclature[P,14]{\(c_F\)}{Annualized operation cost of desalination unit, \$ per t. }
\nomenclature[P,15]{\(\alpha_F\)}{Fresh water-electricity conversion efficiency, MW per t. }
\nomenclature[P,16]{\(F_0\)}{Daily fresh water demand, t. }
\nomenclature[P,17]{\(\eta_l^{ESS}\)}{Efficiency of charging and discharging of the $l$th ESS, dimensionless. }
\nomenclature[P,18]{\(\eta^{TPG}_{k}\)}{Efficiency of the $k$th tidal generation unit, dimensionless. }
\nomenclature[P,19]{\(\tilde{P}_{j,h,d,y}\)}{Nominal generation of the $j$th RES and the $k$th TPG unit given the natural condition at hour $h$, day $d$, year $y$, respectively, MW. }
\nomenclature[P,20]{\(\Gamma^L_y,\Gamma^{TPG}_y\)}{Uncertainty budget of electricity load demand and tidal power generation in year $y$, respectively, dimensionless. }
\nomenclature[P,21]{\(\beta_L,\beta_{TPG}\)}{Deviation coefficient of load demand and tidal height, respectively, dimensionless. }
\nomenclature[P,22]{\(\gamma_t^L,\gamma_t^{TPG}\)}{Uncertainty budget coefficients of load demand and tidal height, respectively, dimensionless. }
\nomenclature[P,23]{\(g\)}{Acceleration due to the Earth's gravity, $m/s^2$. }
\nomenclature[P,24]{\(\rho\)}{Density of sea water, $kg/m^3$. }
\nomenclature[P,25]{\(h^{TPG}_{h,d,y}\)}{Tidal height at hour $h$, day $d$, year $y$, m. }
\nomenclature[P,26]{\(A_k\)}{Area of the $k$th tidal generation unit, $m^2$. }

\nomenclature[V,01]{\(x_i\)}{Binary decision variable of the investment status of the $i$th unit, dimensionless. }
\nomenclature[V,02]{\(P_{i,h,d,y}\)}{Generation of the $i$th device at hour $h$, day $d$, year $y$, MW. }
\nomenclature[V,03]{\(P^{dch}_{l,h,d,y},P^{ch}_{i,h,d,y}\)}{Discharging and charging power of the $l$th energy storage system at hour $h$, day $d$, year $y$, MW. }
\nomenclature[V,04]{\(F_{h,d,y}\)}{Amount of fresh water produced by the desalination unit, t. }
\nomenclature[V,05]{\(SOC_{l,h,d,y}\)}{State of charge of the $i$th energy storage system at hour $h$, day $d$, year $y$, MWh. }
\nomenclature[V,06]{\(LS_{h,d,y}\)}{Load shedding at hour $h$, day $d$, year $y$, MW. }
\nomenclature[V,07]{\(L_{h,d,y}\)}{Electricity load demand at hour $b$, day $h$, year $t$, MW. }
\nomenclature[V,08]{\(\overline{u}^L_{h,d,y}, \underline{u}^L_{h,d,y}\)}{Binary decision variables determining whether load demand at hour $h$, day $d$, year $y$ should be set as its upper or lower bound, respectively, dimensionless. }
\nomenclature[V,09]{\(\overline{u}^{TPG}_{h,d,y}, \underline{u}^{TPG}_{h,d,y}\)}{Binary decision variables determining whether tidal power generation at hour $b$, day $h$, year $t$ should be set as its upper or lower bound, respectively, dimensionless. }

\printnomenclature

\IEEEpeerreviewmaketitle

\section{Introduction}
\IEEEPARstart{M}{icrogrid} is a localized, relatively small-scale power system to provide electricity-related services for customers \cite{shahidehpour2017networked}, its islanded operation is particularly considered when it comes to remote areas, especially the offshore microgrid (OM), for which undersea cables could be expensive and erratic given the huge distance and complicated environment between the island and the mainland. 
Meanwhile, distributed energy resources (DERs) are usually incorporated in the microgrids to serve load demands, reduce electricity costs as well as enhance penetration of renewable energy sources (RESs) \cite{shahidehpour2010role}. 
The DERs are usually classified as dispatchable units (DUs), nondispatchable units (NDUs), energy storage systems (ESSs), etc.
DUs mainly include traditional thermal power generators, which take fossil fuel as an input and could emit toxic gases and generate solid waste, leading to further environmental issues. 
NDUs are mainly RESs, such as wind power and solar energy. 
Both DUs and NDUs are generation units, and ESSs are the units serving as reserves by the charging and discharging process.

Till now studies conducted on dealing with diverse uncertainties in microgrids have been investigated, and the optimization methods such as stochastic optimization (SO), distributionally robust optimization (DRO), and robust optimization (RO) have been considered. 
SO-based microgrid planning methods capture uncertainties based on full information of probabilistic distribution of the underlying uncertainties, while the calculation burden could be huge given the number of scenarios \cite{su2013stochastic, narayan2017risk, wu2011economic, wu2014dynamic, mohammadi2014scenario}. 
When knowing partial information about the probabilistic distribution of uncertainties, DRO \cite{bertsimas2019adaptive} can be applied to deal with uncertainties in power systems \cite{zhao2019two,guevara2020machine,lu2018security,hu2020distributionally}. 
However, given that both SO and DRO are based on probabilistic information of uncertainties, they cannot guarantee the security of the microgrid in some extreme/worst cases, which is obviously necessary for OM.
The RO, as an optimization method that gives the solution under the worst scenario, has been introduced in lots of problems in the area of energy systems \cite{nazari2018application,zhao2013multi,jabr2013robust,rahimiyan2015strategic,zhang2013robust,martinez2013robust}, and is especially a suitable choice for OM given the reliability of its solution. 

Furthermore, several efforts have been made in microgrid planning considering diverse uncertainties using RO. 
In \cite{quashie2018optimal}, a bi-level deterministic planning model (DPM) coupled with reserve capacity is proposed, which was further reformulated as a mathematical program with equilibrium constraints and then transformed into a mixed-integer linear programming problem. 
Literature \cite{zhang2018probability} proposed a probability-weighted RO method, in which uncertainties from both wind power and load demand are captured by probability-weighted uncertainty sets, and the model is solved by a modified column-and-constraint generation (C\&CG) algorithm\cite{zeng2013solving}. 
In \cite{khodaei2014microgrid}, Khodaei et al. considered uncertainties from load demand, RES generation and electricity market prices in an interval-based manner and solved the model with Benders decomposition; differently, uncertainties in \cite{tan2021wind}, including RES generation and ambient temperature are solved with C\&CG. 
Similarly, the model in \cite{wang2014robust} was also solved with C\&CG with the uncertainty set formulated as a polyhedron. 
The nested C\&CG algorithm is utilized in \cite{qiu2018interval} to solve the proposed robust planning model (RPM), where uncertainties from RES generation as well as operating states of the bidirectional converters are modeled with interval-partitioned uncertainty sets that reduce the conservativeness of traditional single-interval uncertainty based robust models. 
In \cite{yang2020interval}, a robust model was established with uncertainties from load demand and RES generation analyzed with the interval analysis method, and the uncertain constraints are converted to deterministic ones so that the model can be solved easily. 



However, none of the above research considered the new elements in OM specifically, such as ocean renewable energy (ORE) including tidal power and wave energy etc, and some special load such as seawater desalination units (SDU). 
ORE should not be ignored for OM planning, given the abundant quantity of ORE on islands, as well as the uniqueness of OM. 
Among all types of ORE, tidal power is one of the most promising. 
Generally, tidal power is caused by periodic changes in sea levels resulting from the gravitational attraction between celestial bodies \cite{frau1993tidal}.
Tidal power has some favorable advantages over traditional RESs, such as wind and solar energy, since it is easier to be predicted with decent accuracy many years in advance \cite{khan2018wave}, making it an appropriate choice for planning. 
Besides, tidal power is expected to have a great potential of $3.6$ TW in 2030 \cite{sleiti2017tidal}, which could be highly helpful in enhancing renewable penetration as well as achieving dual carbon goals. 
Moreover, load of SDU should also be taken into consideration. 
The importance of fresh water for human beings is needless to say, and islands are especially in shortage of fresh water since they are surrounded by ocean. 
For island citizens to get fresh water, two methods are feasible: transporting fresh water from mainland, or desalination. 
Given the distance between mainland and the island, sea water desalination becomes a more economical way. 
Power consumption of SDU is a unique load for offshore microgrid. 
Power consumption of SDU could account for as much as around 17\% of the total predicted electricity load demand, which is a proportion that should not be ignored. 
However, few studies incorporating tidal power into OM planning or taking load demand of SDU into consideration have been conducted till now. 

The main contributions of this paper are summarized as follows: 
\begin{itemize}
  \item [1)] 
A two-stage RPM for OM integrated with SDU and tidal power generation (TPG) is proposed. 
The first stage includes the investment decisions on various DUs and NDUs, as well as ESSs, while in the second stage the operation decisions are also made considering the uncertainties of both load and tidal generation.
  \item [2)]
To analyze the role of TPG in OM accurately, its uncertainty is modeled in terms of tidal height, and scenarios with different tidal delays are also simulated. 
The case studies show the effectiveness of the proposed uncertainty modeling for TPG. 
It is also shown that, TPG could work complementarily with the other generation methods.
  \item [3)]
The C\&CG algorithm is then employed to solve the two-stage RPM. Numerical experiments in different scenarios show the essential role of ORE in OM planning, and the robustness of the planning results.
\end{itemize}

The remainder of the paper is organized as follows. 
Section II gives the framework of the planning problem and method. 
Section III introduces the proposed mathematical model, including the objective functions, the constraints and uncertainty modeling of the OM with tidal power in detail. 
Section IV provides numerical simulations based on the test microgrid cases. 
Section V summarizes the results and concludes the paper. 

\begin{figure}[b]
    \centering
    \includegraphics[width=3in]{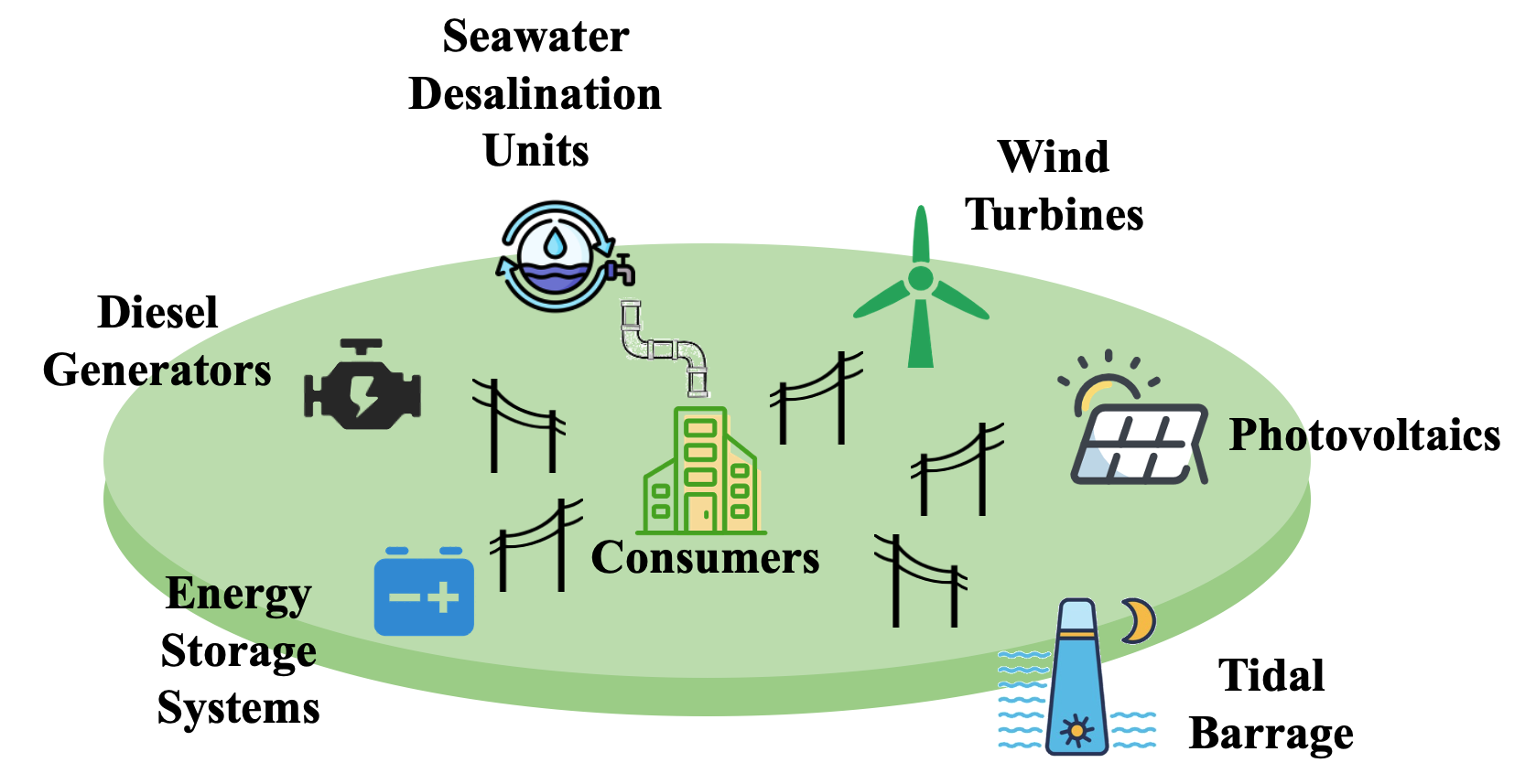}
    \caption{Structure of the OM discussed in this paper.}
    \label{mainland_island}
\end{figure}

\section{Framework of the OM Planning Method}
In this paper, the planning problem of an OM is considered. 
It's assumed that the candidate units of the microgrid include diesel generators (DU), wind turbines, photovoltaics (PV), energy storage systems (ESS) as well as tidal generators, as illustrated in Fig. \ref{mainland_island}. 
DSU, as the only device that is able to produce fresh water, will be invested compulsively and is hence not assumed as one of the candidate units.

\begin{figure}[htbp]
    \centering
    \includegraphics[width=3.3in]{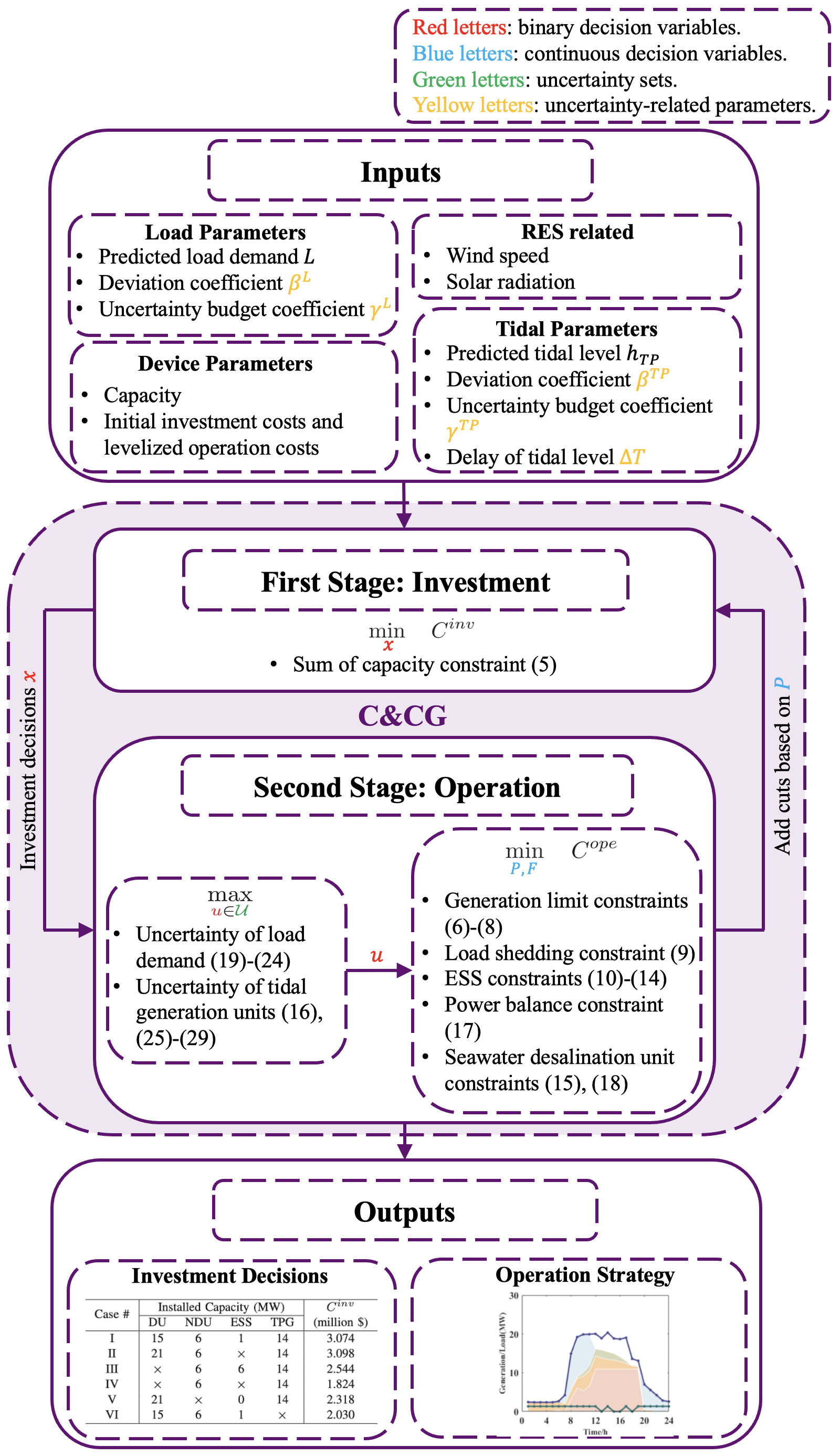}
    \caption{Framework of the planning problem discussed in this paper. }
    \label{structure}
\end{figure}

All the decision variables can be divided into two categories: 

(1) Binary decision variables marked with red in Fig. \ref{structure}, including $x_i,\forall i \in \Omega_{DU}$, $x_j,\forall j \in \Omega_{NDU}$, $x_k,\forall k \in \Omega_{TPG}$, $x_l,\forall l \in \Omega_{ESS}$, for investment decisions of the candidate units, and $\underline{u}_{h,d,y}^L$, $\overline{u}_{h,d,y}^L$, $\underline{u}_{h,d,y}^{TPG}$, $\overline{u}_{h,d,y}^{TPG}$ for uncertainty decisions. 
For simplicity, the aforementioned $x_i$, $x_j$, $x_k$, $x_l$ and $\underline{u}_{h,d,y}^L$, $\overline{u}_{h,d,y}^L$, $\underline{u}_{h,d,y}^{TPG}$, $\overline{u}_{h,d,y}^{TPG}$ are represented by $x$ and $u$ respectively in Fig. \ref{structure} and the rest of the paper;

(2) Continuous decision variables marked with blue in Fig. \ref{structure}, representing power generation quantity of the generation units $P_{i,h,d,y},\forall i\in \Omega_{DU}$, $P_{j,h,d,y},\forall j\in \Omega_{NDU}$, $P_{k,h,d,y},\forall k\in \Omega_{TPG}$, charging and discharging power of the ESSs $P^{ch}_{l,h,d,y},P^{dch}_{l,h,d,y},\forall l\in \Omega_{ESS}$, as well as fresh water production of the SDU $F_{h,d,y}$. 
For simplicity, the aforementioned $P_{i,h,d,y}$, $P_{j,h,d,y}$, $P_{k,h,d,y}$, $P^{ch}_{l,h,d,y}$, $P^{dch}_{l,h,d,y}$ and $F_{h,d,y}$ are represented by $P$ and $F$ in Fig. \ref{structure} and the rest of the paper. 

The inputs of the proposed OM planning model comprise the predicted data of load demand, wind speed, solar radiation and tidal height, the device parameters, and the parameters related to uncertainty modeling of deviation coefficients, uncertainty budget coefficients, delay of tidal height, etc. 

The solution procedure is composed of two stages. 
Investment decisions are made in the first stage, also known as the planning stage. 
The second stage is known as the operation stage, where the operation strategy in the worst scenario could be derived from the investment decisions determined in the first stage. 
If investment decisions from the first stage bring infeasibility to the second stage, cuts will be formed and returned to the first stage as additional constraints, where the investment decisions will be optimized again until the optimization problem in the second stage is feasible. 
Therefore, the outputs of the model include first-stage investment decisions and second-stage operation strategies.
The framework of the planning-operation co-optimization model discussed in this paper is illustrated in Fig. \ref{structure}. 

\section{Model Formulation}
In this section, a DPM for the OM, i.e., a model without consideration of uncertainties, is proposed first. 
Uncertainties are then modeled and the uncertainty sets are formulated, which are then integrated into the original model to derive the robust model. 

\subsection{Deterministic Planning Model}
\subsubsection{Objective Function}
The objective function \eqref{obj_two_terms} of the DPM includes investment cost $C^{inv}$ for capital expenditure of the devices, and operation cost $C^{ope}$ for the real-time scheduling of the microgrid. 
The object is to minimize the total cost to make the most economical decisions. 
\begin{equation}
    \label{obj_two_terms}
    \min \quad C^{total}=C^{inv}+C^{ope}
\end{equation}
\begin{equation}
\label{C^inv}
    C^{inv}=\sum_{y} \kappa_y \left[ \begin{matrix}
 \sum_{i\in\Omega_{DU}} cc_{i} RC_i x_i\\
 +\sum_{j\in\Omega_{NDU}} cc_{j} RC_j x_j\\
 +\sum_{k\in\Omega_{TPG}} cc_{k} RC_k x_k\\
+\sum_{l\in\Omega_{ESS}}\left(cp_{l} RP_l+ce_{l} RC_l\right) x_l
\end{matrix} \right]
\end{equation}
where the first three terms in \eqref{C^inv} stand for investment costs of DUs, NDUs and TPG units, and the fourth term stands for investment costs of the ESSs. 
$x$ is the binary decision variable for the device, $x=1$ indicates that the device is selected to be invested, $x=0$ otherwise. 
$cc$ is the annualized investment cost of the device per MW, and $RC$ is the rated capacity of the corresponding device. 
The investment cost of the $l$th ESS is formulated as the combination of the cost for rated power $RP_l$ and rated capacity $RC_l$ (here unit: MWh). 

Note that there is a coefficient of the present-worth value of the $y$th year $\kappa_y$ in $C^{inv}$ so as to count for the effect of the discount rate $dr$. 
$\kappa_y$ can be calculated based on discount rate $dr$ as follows: 
\begin{equation} \label{kappa_t}
    \kappa_y=\frac{1}{\left(1+dr\right)^{y-1}},\forall y
\end{equation}

The total operation cost $C^{ope}$ is formed as follows: 
\begin{equation}
\label{C^ope}
C^{ope}=\sum_{y}\sum_{d}\sum_{h}
\left[
\begin{matrix}
    \sum_{i\in\Omega_{DU}}c_i P_{i,h,d,y}\\
    +\sum_{k\in\Omega_{TPG}}c_k P_{k,h,d,y}\\
    +c_F F_{h,d,y} +\nu LS_{h,d,y}
\end{matrix}
\right],\forall h,d,y 
\end{equation}
where the first two terms represent the operation costs of DUs and TPG, calculated as the product of the levelized operation cost $c$ and the real-time operation power $P_{h,d,y}$. 
The third term is the operation cost of SDU, where $c_F$ is the levelized operation cost of SDU, i.e., the operation cost of producing each ton of fresh water, and $F_{h,d,y}$ is the amount of fresh water produced at each time period. 
The last term is punishment for load shedding, calculated as the product of the punishment factor $\nu_{h,d,y}$ and the shedding-load power $LS_{h,d,y}$.
The punishment factor $\nu$ for load shedding, also known as the value of lost load, is set as a sufficiently large number so as to drive load shedding to zero. 
\subsubsection{Constraints}
The constraints considered in the DPM of the OM planning problem are as follows: 
\begin{equation}
\label{deter_cons_inv_1}
    L^{\max}\leq \sum_{i}RP_ix_i+\sum_{j}RP_jx_j+\sum_{k}RP_k x_k
\end{equation}
\begin{equation}
\label{deter_cons_ope_2}
    0\leq P_{i,h,d,y}\leq RP_i x_i
\end{equation}
\begin{equation} \label{deter_cons_ope_3}
    0\leq P_{j,h,d,y}\leq RP_j x_j
\end{equation}
\begin{equation} \label{deter_cons_ope_4}
    0\leq P_{k,h,d,y}\leq RP_k x_k
\end{equation}
\begin{equation} \label{deter_cons_ope_5}
    0\leq LS_{h,d,y}\leq L_{h,d,y}
\end{equation}
\begin{equation} \label{deter_cons_ope_6}
    0 \leq P_{l,h,d,y}^{dch} \leq RP_l  x_l
\end{equation}
\begin{equation} \label{deter_cons_ope_7}
    0 \leq P_{l,h,d,y}^{ch} \leq RP_l  x_l 
\end{equation}
\begin{equation} \label{deter_cons_ope_9}
    \begin{aligned}
        SOC_{l, h+1, d, y}&=SOC_{l,h,d,y}+P_{l,h,d,y}^{ch} \eta_l^{ESS}-P_{l,h,d,y}^{dch} / \eta_l^{ESS}\\
    \end{aligned} 
\end{equation}
\begin{equation} \label{deter_cons_ope_10}
    0 \leq SOC_{l,h,d,y} \leq RC_l x_l
\end{equation}
\begin{equation} \label{deter_cons_ope_11}
    SOC_{l,1,d,y}=SOC_{l,24,d,y}
\end{equation}
\begin{equation} \label{deter_cons_ope_12}
    0 \leq F_{h,d,y} \leq RC_F
\end{equation}
\begin{equation} \label{tidal_generation_calculation}
    \tilde{P}^{TPG}_{k,h,d,y}=\frac{1}{2}\cdot \rho \cdot g \cdot {h_{h,d,y}^{TPG}}^2 \cdot A_k \cdot \eta^{TPG}_k/3600,\quad \forall k\in \Omega_{TPG}
\end{equation}
\begin{equation} \label{deter_cons_ope_1}
    \begin{aligned}
        &\sum_{i}P_{i,h,d,y}+\sum_{j}P_{j,h,d,y}+\sum_{k}P_{k,h,d,y}+\sum_{l}\left(P_{l,h,d,y}^{dch}-P_{l,h,d,y}^{ch}\right)\\
        &=L_{h,d,y}+\alpha_F F_{h,d,y}-LS_{h,d,y}
    \end{aligned}
\end{equation}
\begin{equation} \label{deter_cons_ope_ds_sum}
    \sum_{h} F_{h,d,y}\geq F_0
\end{equation}
\[\begin{array}{c}
\quad \forall i\in \Omega_{DU},\quad \forall j\in \Omega_{NDU},\quad \forall k\in \Omega_{TPG},\quad \forall l \in \Omega_{ESS},\\
\forall h,d,y
\end{array}\]

All the $x$ are binary decision variables of investment states of the DERs, $x=1$ indicates the device should be invested and $x=0$ otherwise. 
Constraint \eqref{deter_cons_inv_1} guarantees the sum of the capacities of all the installed units is higher than the maximal predicted load demand $L^{\max}$ to guarantee the feasibility of the planning problem. 
Constraint \eqref{deter_cons_ope_2} is a capacity limit constraint, ensuring the generation of each DU is no greater than the rated power of the corresponding unit when installed, and is exactly $0$ when not installed. 
The next two constraints, \eqref{deter_cons_ope_3} and \eqref{deter_cons_ope_4} are formulated in a similar manner, imposing restrictions on the generation of RESs including wind power, solar energy and tidal power. 
Constraint \eqref{deter_cons_ope_5} is to limit load shedding $LS_{h,d,y}$ in the range between $0$ and load demand $L_{h,d,y}$ at the moment to guarantee the load shedding is no higher than the load demand. 
Charging power $P^{ch}_{l,h,d,y}$ and discharging power $P^{dch}_{l,h,d,y}$ of the ESSs are restricted by \eqref{deter_cons_ope_6} and \eqref{deter_cons_ope_7} in the range between $0$ and $RP_{l}$ of the corresponding unit. 
Equality constraint \eqref{deter_cons_ope_9} is utilized to calculate the state of charge (SOC) of the ESSs. 
The SOC of each ESS at each time period $SOC_{l,h+1,d,y}$ is calculated based on the last time period $SOC_{l,h,d,y}$ by appending a term related to the charging power of the last time period $P^{ch}_{l,h,d,y}$ and subtracting a term related to the discharging power of the last time period $P^{dch}_{l,h,d,y}$. 
Note that efficiencies of the ESSs $\eta_l^{ESS}$ are considered here to account for the fact that not all the power being charged or discharged can be received on the other end perfectly. 
In \eqref{tidal_generation_calculation}, generation of a tidal barrage in an hour can be calculated from tidal height \cite{lamb1994hydrodynamics}, where $\rho$ is density of sea water, $g$ is acceleration due to the Earth's gravity, $h_{TPG}$ is tidal height, $A_i$ and $\eta^{TPG}_{k}$ are the area and efficiency of the $k$th tidal generation unit, respectively. 
SOC of each ESS is restricted within the range of its rated capacity by \eqref{deter_cons_ope_10}, and the SOC of each device at the start of the day $SOC_{l,h,d,y}$ is required to be the same as the value at the end of the day $SOC_{l,H,d,y}$ as required by \eqref{deter_cons_ope_11}, which is helpful to extend battery life, and helps to ensure that the ESSs can always be utilized when being needed. 
Power balance is guaranteed in \eqref{deter_cons_ope_1}, where sum of generation of all the generating units is required to be equal to sum of predicted load demand and load demand from SDU, denoted by $\alpha_F F_{h,d,y}$, where $\alpha_F$ is the fresh water-electricity conversion efficiency, i.e., the amount of power needed to produce each ton of fresh water. 
Efficiencies of the ESSs are not included here since they are already considered when calculating $SOC$. 
In \eqref{deter_cons_ope_ds_sum}, fresh water production in each day is required to meet the daily fresh water demand $F_0$.
Since the storage of fresh water is generally large and simple \cite{WANG2020105707}, fresh water demand balance is considered on a daily basis, rather than in each hour. 

To conclude, the DPM of the OM can be summarized in the following compact form: 
\begin{equation*}\label{deterministic_model}
    \begin{aligned}
        & \min_{x,P,LS,F} \quad C^{total} \\
        & s.t. \quad \text{Constraints }\eqref{deter_cons_inv_1}-\eqref{deter_cons_ope_ds_sum}.
\end{aligned}
\end{equation*}

\subsection{Uncertainty Characterization in Offshore Microgrid}
Uncertainties from load demand and tidal generation are formulated in this part. 

\subsubsection{Load Demand Uncertainty Modeling}
The uncertainty set of load demand $\mathcal{U}_L$ is formulated as follows: 
\begin{equation}
    \label{uncertainty_set_load_1}
    \mathcal{U}_L:L_{h,d,y} =\tilde{L}_{h,d,y}-\underline{L}_{h,d,y} \underline{u}_{h,d,y}^L+\overline{L}_{h,d,y} \overline{u}_{h,d,y}^L
\end{equation}
\begin{equation} \label{uncertainty_set_load_2}
\underline{u}_{h,d,y}^L+\overline{u}_{h,d,y}^L \leq 1
\end{equation}
\begin{equation} \label{uncertainty_set_load_4}
\overline{L}_{h,d,y}=\underline{L}_{h,d,y}=\beta_L \tilde{L}_{h,d,y}
\end{equation}
\begin{equation} \label{uncertainty_set_load_3}
    \sum_{d} \sum_{h} \left(\underline{u}_{h,d,y}^L+\overline{u}_{h,d,y}^L\right)\leq \Gamma^L_y
\end{equation}
\begin{equation} \label{uncertainty_set_load_6}
    \Gamma^L_y=\gamma_y^L \cdot D \cdot H
\end{equation}
\[\begin{array}{c}
\forall h,d,y
\end{array}\]
where $\overline{u}_{h,d,y}^L$ and $\underline{u}_{h,d,y}^L$ are binary decision variables indicating whether the load is increased to its upper limit or decreased to its lower limit, respectively. 
Constraint \eqref{uncertainty_set_load_2} guarantees the two binary decision variables won't be set as $1$ at the same time.
$\overline{L}_{h,d,y}$ and $\underline{L}_{h,d,y}$ are the upper and lower deviation of the uncertainty set, respectively, and can be calculated by \eqref{uncertainty_set_load_4}. 
In \eqref{uncertainty_set_load_4}, $\beta_L$ is the deviation coefficient of load demand, and is a positive constant less than $1$. 
Constraint \eqref{uncertainty_set_load_1} can be further elaborated as follows: 
\begin{equation} \label{uncertainty_set_load_1_elaborated}
    L_{h,d,y}=\left \{ 
\begin{aligned}
    & \tilde{L}_{h,d,y}-\underline{L}_{h,d,y}\text{, if }\underline{u}_{h,d,y}^L=1,\overline{u}_{h,d,y}^L=0\\
    & \tilde{L}_{h,d,y}+\overline{L}_{h,d,y}\text{, if }\underline{u}_{h,d,y}^L=0,\overline{u}_{h,d,y}^L=1\\
    & \tilde{L}_{h,d,y}\text{, if }\underline{u}_{h,d,y}^L=0,\overline{u}_{h,d,y}^L=0
\end{aligned}
    \right.
\end{equation}
hence when $\underline{u}_{h,d,y}^L=1$ and $\overline{u}_{h,d,y}^L=0$, a lower deviation is subtracted from the predicted value, and when $\overline{u}_{h,d,y}^L=1$ and $\underline{u}_{h,d,y}^L=0$, an upper deviation is added to the predicted value. 
When both of them are $0$, the predicted value of load demand is achieved. 
Constraint \eqref{uncertainty_set_load_3} sets an uncertainty budget $\Gamma^L_y$ for the uncertainty set \cite{bertsimas2004price}, restricting the number of deviations of all time periods in one year in which the load demand is far away from its predicted value, hence this parameter helps to adjust the conservatism of the model. 
$\Gamma^L_y$ can be calculated from \eqref{uncertainty_set_load_6}, and it is the coefficient controlling the percentage of the deviated scenarios, and $\gamma_y^L$ is within the range of $\left[0,1\right]$. 
For instance, when $\gamma_y^L=\Gamma_y^L=0$, load demands in all time periods are assumed as the corresponding predicted values, thus, there is no robustness considered, and the model is deterministic. 
When setting $\gamma_y^L$ as a higher value, $\Gamma_y^L$ also goes up and a higher degree of robustness is achieved and the model is more conservative. 
Adjusting the uncertainty budget helps to adjust the extent to which uncertainty is considered. 
A higher uncertainty budget allows more deviation points, a higher degree of uncertainty is thus achieved and vice versa.

\subsubsection{Tidal Power Uncertainty Modeling}
\begin{figure}[!hbt]
    \centering
    \includegraphics[width=2.6in]{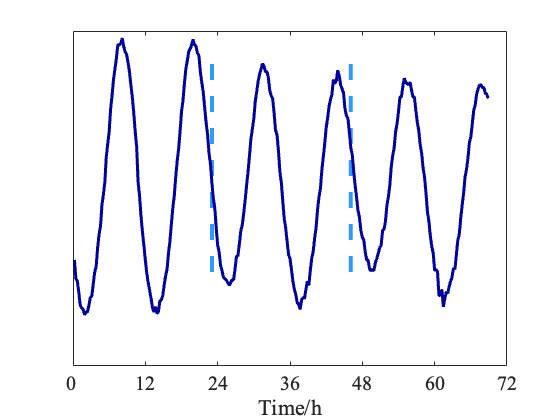}
    \caption{Typical tidal heights in consecutively three days. }
    \label{tidal_peaks}
\end{figure}

Tides are generally semidiurnal, with two high tides and two low tides per day, with two high tides located at 4:00-11:00 am and 4:00-11:00 pm, respectively, as shown in Fig. \ref{tidal_peaks}. 

Since the effect of tidal heights can be reflected in TPG as in \eqref{tidal_generation_calculation}, uncertainty of tidal heights is formed as an interval-based set of TPG in a similar manner as follows: 
\begin{equation} \label{uncertainty_set_tidal_1}
    \mathcal{U}_{TPG}:P^{TPG}_{k,h,d,y}=\tilde{P}^{TPG}_{k,h,d,y}-\underline{P}^{TPG}_{k,h,d,y}\underline{u}_{k,h,d,y}^{TPG}+\overline{P}^{TPG}_{k,h,d,y}\overline{u}_{k,h,d,y}^{TPG}
\end{equation}
\begin{equation} \label{uncertainty_set_tidal_2}
    \underline{u}_{k,h,d,y}^{TPG}+\overline{u}_{k,h,d,y}^{TPG} \leq 1 
\end{equation}
\begin{equation} \label{uncertainty_set_tidal_4}
    \overline{P}^{TPG}_{k,h,d,y}=\underline{P}^{TPG}_{k,h,d,y}=\beta_{TPG} \tilde{P}^{TPG}_{k,h,d,y}
\end{equation}
\begin{equation} \label{uncertainty_set_tidal_3}
    \sum_{d} \sum_{h} \left(\underline{u}_{k,h,d,y}^{TPG}+\overline{u}_{k,h,d,y}^{TPG}\right)\leq \Gamma^{TPG}_y
\end{equation}
\begin{equation} \label{uncertainty_set_tidal_6}
    \Gamma^{TPG}_y=\gamma_y^{TPG} \cdot D \cdot H
\end{equation}
\[\begin{array}{c}
\forall k\in \Omega_{TPG}, \forall h,d,y
\end{array}\]

\begin{figure}[!hbt]
    \centering
    \includegraphics[width=2.1in]{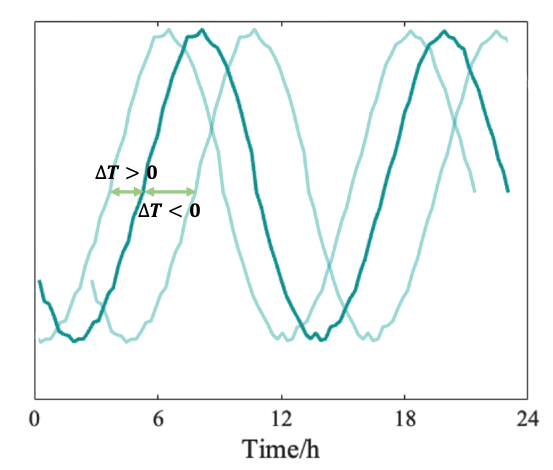}
    \caption{Modeling of uncertainty time of peaks of tidal heights. }
    \label{tidal_time_uncertainty}
\end{figure}

Besides, time of peak of tidal level is also assumed as uncertain and the time delay $\Delta T$ is assumed to be $\pm 4$ hours within the predicted time. 
$\Delta T > 0$ corresponds to the case of the peaks coming earlier and $\Delta T < 0$ otherwise. 
When $\Delta T = 0$, the predicted time of peaks is assumed to be achieved. 
When the predicted levels are moved forward or backward, the vacated points are filled with 0, as illustrated in Fig. \ref{tidal_time_uncertainty}. 
Scenarios with different $\Delta T$ are simulated and analyzed in the case studies. 

\subsection{Robust Planning Model for Offshore Microgrid}
A two-stage RPM can be established considering the uncertainties of load demand and tidal generation. 
In the first stage, the investment status of the units are determined by a master problem to minimize the investment cost, assuming load demand and tidal generation follow the predicted values. 
After achieving the worst case, the operation strategies are determined in the second stage of the model to minimize the operation cost given the worst case. 
The derived model is as follows: 
\begin{equation*}\label{robust_model}
    \begin{aligned}
        & \min_{x} \quad C^{inv}+\max_{u}\min_{P,LS,F} \quad C^{ope} \\
        & s.t. \quad \text{Constraints } \eqref{deter_cons_inv_1} -\eqref{uncertainty_set_load_6},\eqref{tidal_generation_calculation}-\eqref{uncertainty_set_tidal_6}.
    \end{aligned}
\end{equation*}
where $\mathcal{U}=\mathcal{U}_L \cup \mathcal{U}_{TPG}$ is the union of the considered uncertainty sets. The two-stage RPM is solved using C\&CG algorithm \cite{zeng2013solving}, which shows better computational efficiency in terms of convergence speed and time compared with Benders Decomposition \cite{zeng2013solving,zhao2012robust}. 

\section{Case Studies}
Numerical cases are tested in this section. 
The basic parameters of the test system are introduced first, followed by the results and analyses of the numerical simulations under different conditions. 

The candidate devices include six DUs, two NDUs, three ESSs and four TPG units. 
The fixed parameters of the candidate units are listed in TABLEs \ref{table_dispatchable}-\ref{table_TP}. 
The efficiencies of charging and discharging of the ESSs are all assumed as 90\%, and the planning horizon is 20 years. 
Daily fresh water demand is set as 9000 t, and the capacity of SDU is set as 450 t/h. 
Annualized investment costs and levelized operation costs of the SDU are set as 1.8M \$ and 1 \$/t. 
Fresh water-electricity conversion efficiency $\alpha_F$ is 3 kW/t, namely, each ton of fresh water production consumes 3 kW of electricity. 

To start with, six DER configuration strategies are designed in TABLE \ref{table_case3_list}, where $\checkmark$ and $\times$ stand for the corresponding device is allowed and not allowed to be invested, respectively.

\begin{table}[!t]
\renewcommand{\arraystretch}{1.1}
\caption{Parameters of the Candidate DUs}
\label{table_dispatchable}
\centering
\begin{tabular}{cccc}
\hline
Unit No. & Capacity (MW) & \makecell[c]{Levelized\\Operation\\Cost (\$/MWh)} & \makecell[c]{Annualized\\Investment\\Cost (\$/MW)}\\
\hline
1 & 6 & 140 & 44,000\\
2 & 5 & 130 & 54,000\\
3 & 4 & 120 & 64,000\\
4 & 3 & 110 & 74,000\\
5 & 2 & 100 & 84,000\\
6 & 1 & 90 & 94,000\\
\hline
\end{tabular}
\end{table}

\begin{table}[!t]
\renewcommand{\arraystretch}{1.1}
\caption{Parameters of the Candidate NDUs}
\label{table_nondispatchable}
\centering
\begin{tabular}{ccccc}
\hline
\makecell[c]{Unit\\No.} & \makecell[c]{Capacity\\(MW)} & \makecell[c]{Levelized\\Operation\\Cost (\$/MWh)} & \makecell[c]{Annualized\\Investment\\Cost (\$/MW)} & \makecell[c]{Type of\\Energy}\\
\hline
1 & 4 & - & 150,000 & WT\\
2 & 2 & - & 90,000 & PV\\
\hline
\end{tabular}
\end{table}

\begin{table}[!t]
\renewcommand{\arraystretch}{1.1}
\caption{Parameters of the Candidate ESSs}
\label{table_ESS}
\centering
\begin{tabular}{ccccc}
\hline
Unit No. & \makecell[c]{Rated\\Power\\(MW)} & \makecell[c]{Rated\\Energy\\(MWh)} & \makecell[c]{Annualized\\Investment\\Cost - Power\\(\$/MW)} & \makecell[c]{Annualized\\Investment\\Cost - Energy\\(\$/MWh)}\\
\hline
1 & 1 & 6 & 60,000 & 30,000\\
2 & 2 & 6 & 30,000 & 30,000\\
3 & 3 & 6 & 20,000 & 30,000\\
\hline
\end{tabular}
\end{table}

\begin{table}[!t]
\renewcommand{\arraystretch}{1.1}
\caption{Parameters of the Candidate TPG Units}
\label{table_TP}
\centering
\begin{tabular}{cccc}
\hline
Unit No. & Capacity (MW) & \makecell[c]{Levelized\\Operation\\Cost (\$/MWh)} & \makecell[c]{Annualized\\Investment\\Cost (\$/MW)}\\
\hline
1 & 5 & - & 54,000\\
2 & 4 & - & 72,000\\
3 & 3 & - & 90,000\\
4 & 2 & - & 108,000\\
\hline
\end{tabular}
\end{table}

\begin{table}[!t]
\renewcommand{\arraystretch}{1.1}
\caption{Device Configuration Strategies of Test Cases}
\label{table_case3_list}
\centering
\begin{tabular}{ccccc}
\hline
Case \# & DU & NDU & ESS & TPG\\
\hline
I & \checkmark & \checkmark & \checkmark & \checkmark \\
II & \checkmark & \checkmark & $\times$ & \checkmark \\
III & $\times$ & \checkmark & \checkmark & \checkmark \\
IV & $\times$ & \checkmark & $\times$ & \checkmark \\
V & \checkmark & $\times$ & \checkmark & \checkmark \\
VI & \checkmark & \checkmark & \checkmark & $\times$ \\
\hline
\end{tabular}
\end{table}

\subsection{Analysis of Effectiveness of Robust Planning} 
\begin{table}[!bt]
\renewcommand{\arraystretch}{1.1}
\caption{Comparison of Investment Decisions and Costs\\between RPM and DPM under Different DER Configurations}
\label{table_result_decision_case0_sufa}
\centering
\begin{tabular}{cc|cccc|c|c}
\hline
\multicolumn{2}{c|}{\multirow{2}{*}{Case \#}} & \multicolumn{4}{c|}{Installed Capacity (MW)} & \multirow{2}{*}{\makecell[c]{$C^{inv}$\\(M \$)}} & \multirow{2}{*}{\makecell[c]{$LS$\\(MW)}} \\ \cline{3-6}
~ & ~ & DU & NDU & ESS & TPG & ~ & ~ \\ \hline
\multirow{2}{*}{I} & DPM & 15 & 6 & 5 & 14 & 3.314 & \multirow{2}{*}{37.47} \\
~ & RPM & 21 & 6 & 5 & 14 & 3.578 & ~ \\ \hline
\multirow{2}{*}{II} & DPM & 15 & 6 & - & 14 & 2.834 & \multirow{2}{*}{49.31} \\
~ & RPM & 21 & 6 & - & 14 & 3.098 & ~ \\ \hline
\multirow{2}{*}{V} & DPM & 15 & - & 6 & 14 & 2.774 & \multirow{2}{*}{66.04} \\
~ & RPM & 21 & - & 6 & 14 & 3.038 & ~ \\ \hline
\multirow{2}{*}{VI} & DPM & 15 & 6 & 3 & $\times$ & 2.030 & \multirow{2}{*}{61.51} \\
~ & RPM & 21 & 6 & 5 & - & 2.534 & ~ \\
\hline
\end{tabular}
\end{table}
In this section, investment decisions and the corresponding costs of RPM is compared with those of DPM to verify the robustness. 
Uncertainty budget coefficient $\gamma^L$ and deviation coefficient of load demand $\beta^L$ are both set as $0.5$ in the RPM. 
Several scenarios representing different weather and load conditions are simulated. 
Investment decisions and costs of two typical days are listed in TABLE \ref{table_result_decision_case0_sufa}. 
Typical day I has lower electricity load demand and less solar radiation, which can be regarded as fall, and typical day II has relatively higher electricity load demand and more solar radiation, which can be regarded as summer. 
The last column in TABLE \ref{table_result_decision_case0_sufa} show the quantity of load shedding when implementing the investment decisions from DPM into RPM in the same case. 
Quantities of load shedding of the opposite scenarios are not listed since there is definitely no load shedding given the more abundant devices determined in the RPM. 

It is obvious that $C^{inv}$ of RPM under the discussed cases are always higher than those of DPM. 
Nevertheless, as shown in TABLE \ref{table_result_decision_case0_sufa}, there are more devices invested in all the discussed cases, hence robustness is guaranteed by more devices and more investment. 
Besides, the installed capacity of DU in all the four discussed cases tends to increase from DPM to RPM, demonstrating DUs' indispensable role in handling uncertainty and enhancing robustness. 
Also, when implementing investment decisions from DPM into RPM, load demands are curtailed by some amount since the decisions from DPM are not robust enough, being unable to handle the worst scenario in the uncertainty sets, demonstrating the robustness of the RPM. 

\subsection{Analysis of Uncertainty of Load Demand} 
\begin{figure}[!hbt]
    \centering
    \includegraphics[width=2.6in]{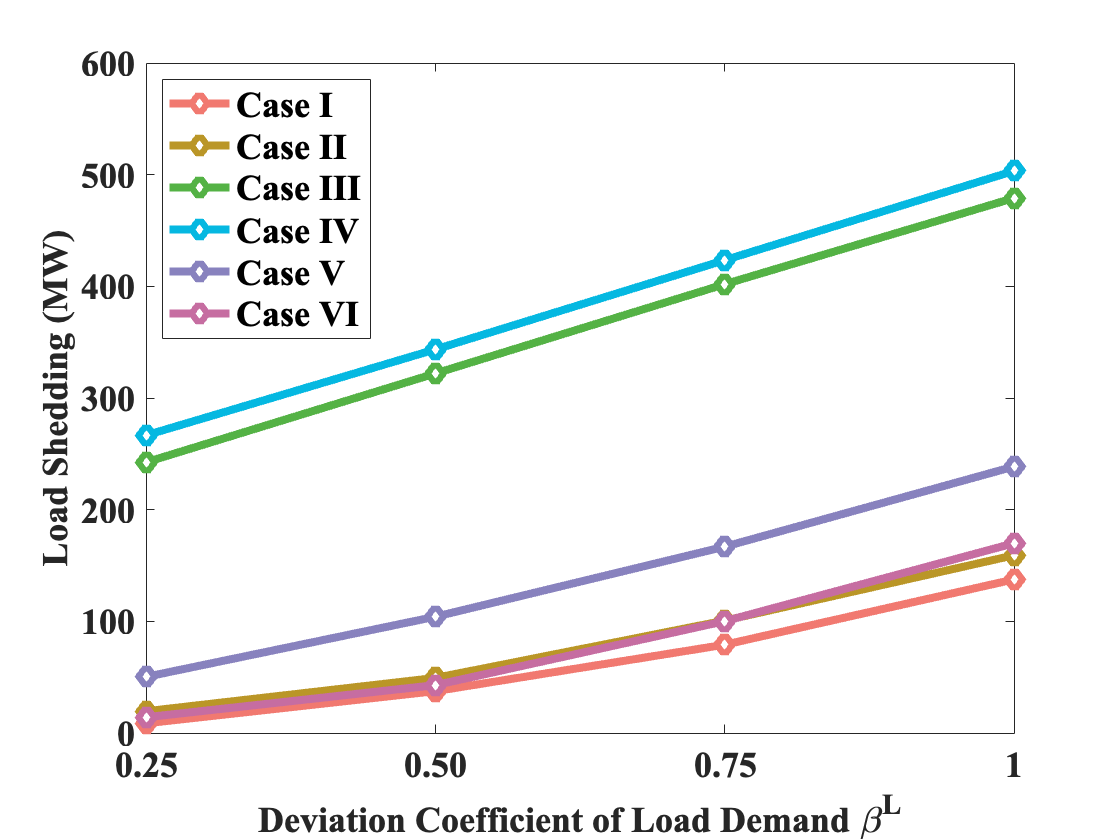}
    \caption{Load shedding of the RPM with different $\beta^L$ given investment decisions of the DPM. }
    \label{case1_LS}
\end{figure}

\begin{figure*}[!hbt]
    \centering
    \includegraphics[width=6.5in]{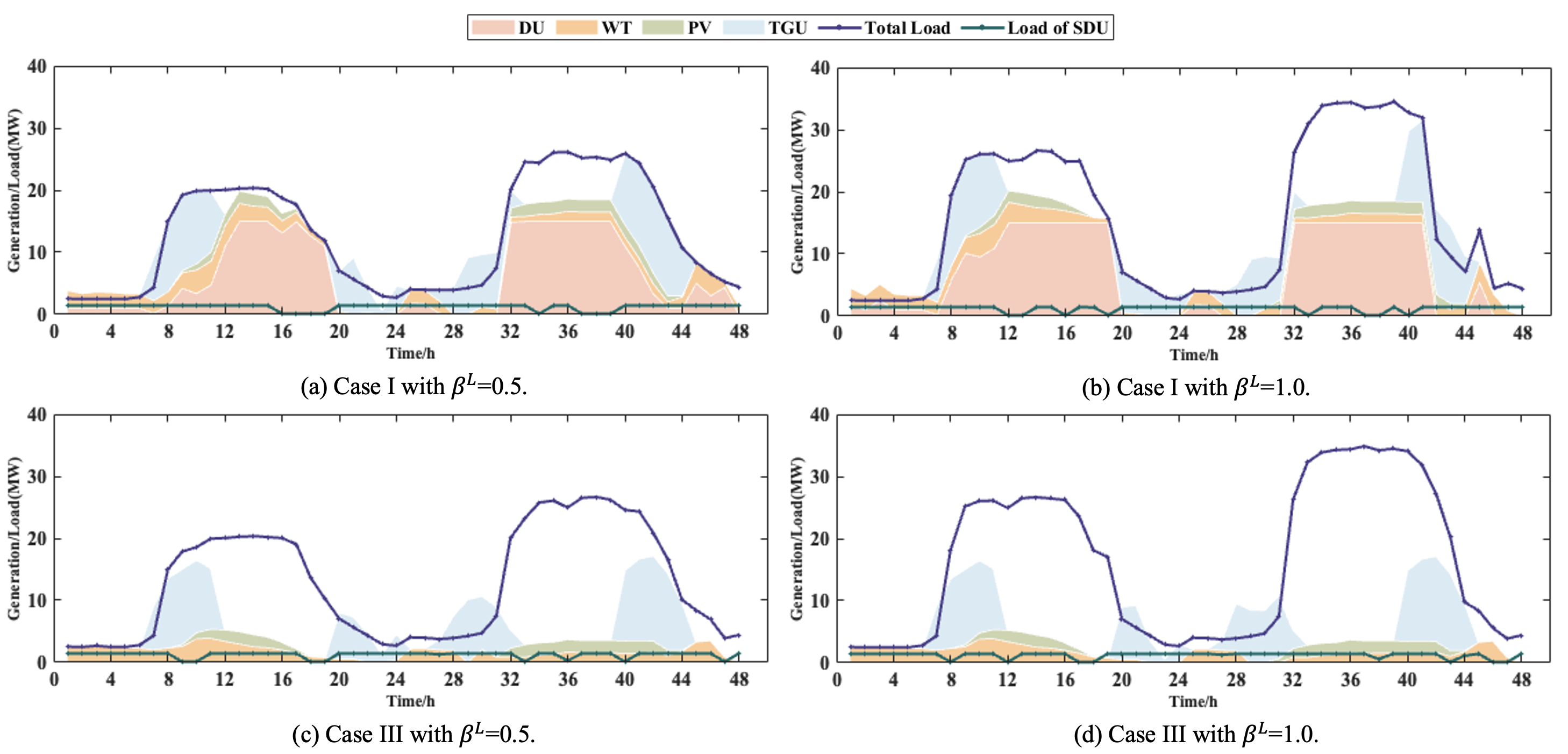}
    \caption{Scheduling diagrams of cases I and IV with $\gamma^L$ set as 0.5 and $\beta^L$ set as 0.5 or 1 in scenario I. }
    \label{case1_operation_figures}
\end{figure*}

In this part, effect of uncertainty from load demand on load shedding is analyzed. 
To start with, all the uncertainty-related coefficients are set as $0$ under the DER configurations of cases I-VI and the investment decisions in these cases are derived and listed in TABLE \ref{table_result_decision_case1_spfa}. 
The investment decisions are then integrated into the system with $\beta^L$ varying from $0.25$ to $1$ and $\gamma^L$ set as $0.5$. 
The resulting load shedding are plotted in Fig. \ref{case1_LS}, and scheduling diagrams of cases I and IV with $\gamma^L$ set as 0.25 or 0.5 are shown in Fig. \ref{case1_operation_figures}, where the white unfilled parts represent the curtailed load demand. 

\begin{table}[bp]
\renewcommand{\arraystretch}{1.1}
\caption{Investment Decisions and Costs of DPM under Different DER Configurations}
\label{table_result_decision_case1_spfa}
\centering
\begin{tabular}{c|cccc|c}
\hline
\multirow{2}{*}{Case \#} & \multicolumn{4}{c|}{Installed Capacity (MW)} & $C^{inv}$\\ \cline{2-5}
~ & DU & NDU & ESS & TPG & (M \$)\\ \hline
I & 15 & 6 & 5 & 14 & 3.314 \\
II & 15 & 6 & - & 14 & 2.834 \\
III & - & 6 & 6 & 14 & 2.544 \\
IV & - & 6 & - & 14 & 1.824 \\
V & 15 & - & 6 & 14 & 2.774 \\
VI & 15 & 6 & 3 & - & 2.030 \\
\hline
\end{tabular}
\end{table}

There is an obvious increasing trend in Fig. \ref{case1_LS} as the deviation coefficient of load demand $\beta^L$ increases from 0.25 to 1. 
Such a trend is intuitive since a higher $\beta^L$ leads to higher load demand to be satisfied with consideration of uncertainty, and given the limited ability of the invested devices derived from the deterministic cases, more loads has to be curtailed. 
Also, load in cases III and IV, i.e., the two cases without investment of DUs, is curtailed by a relatively large quantity as can be found in Fig. \ref{case1_LS} as well as Fig. \ref{case1_operation_figures}(c)(d), demonstrating DUs' significant and reliable role in meeting load demands. 
By comparing cases III and IV, it is easy to discover that the load shedding of case IV is slightly higher, showing ESSs' contribution in helping to satisfy the loads. 
However, the contribution is not that significant since ESSs' main function lies in smoothing fluctuation of load demand, but not satisfy the load directly. 

It is worth mentioning that SDUs mainly work during the early morning and late evening, just complementary to the time domain distribution of load demand, thus helping ease the load pressure during the peak time and make full use of generation during the other periods of the day. 
Part of the fresh water is produced during the peaks, when the generation from solar energy is relatively abundant. 
Such characteristics of the SDUs make full use of lots of the generation methods throughout the day. 

\begin{figure*}[bp]
    \centering
    \includegraphics[width=6.5in]{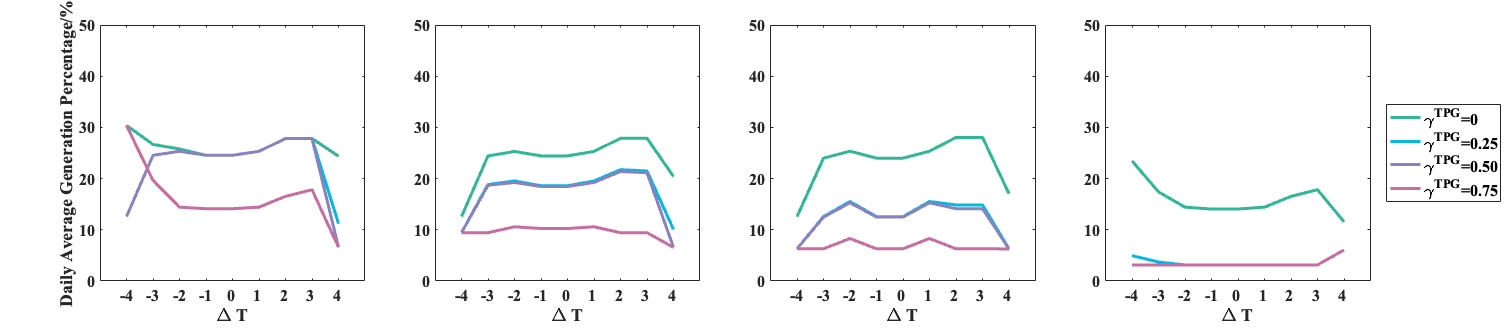}
    \caption{Daily average generation percentage from TPG in OM with different values of $\beta^{TPG}$, $\gamma^{TPG}$ and $\Delta T$. }
    \label{case2_curves_sufa}
\end{figure*}

\begin{figure*}[htbp]
    \centering
    \includegraphics[width=6.5in]{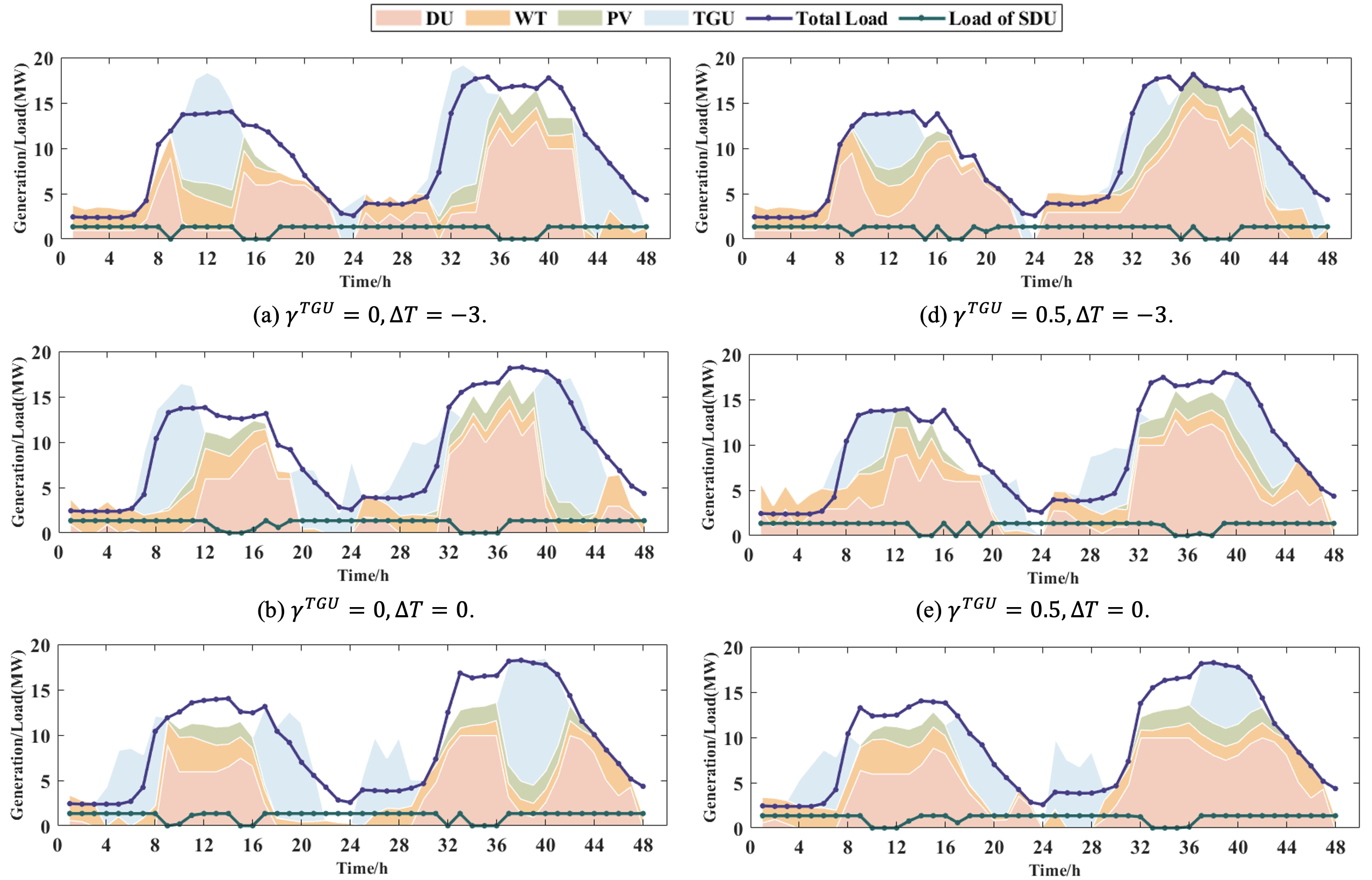}
    \caption{Scheduling diagrams with $\beta^{TPG}$ set as $0.5$, $\Delta T$ set as $-3$, $0$ or $+3$ and $\gamma^{TPG}$ set as $0$ or $0.5$. }
    \label{case2_operation_figures}
\end{figure*}

\subsection{Analysis of Uncertainty of Tidal Generation} 
In this section, uncertainties of tidal height and time of tidal peak are analyzed. 
Different values of $\Delta T$, $\gamma^{TPG}$ and $\beta^{TPG}$ are simulated and the daily average generation percentage of the simulated scenarios, as shown in Fig. \ref{case2_curves_sufa}, are compared and analyzed. 
The daily scheduling results are shown in Fig. \ref{case2_operation_figures}. 

It can be discovered that, generally speaking, a higher $\gamma^{TPG}$ leads to lower generation percentage from TPG, and a higher $\beta^{TPG}$ shows similar effects. 
That is because both higher $\gamma^{TPG}$ and $\beta^{TPG}$ indicate higher level of uncertainty of TPG, and the worst cases given such parameters lead to lower tidal levels, and thus less power generation from the TPG units, which further lead to lower daily average generation percentage. 
As can be observed in Fig. \ref{case2_curves_sufa}, due to the high level of uncertainty, there is sometimes even no TPG units invested when $\gamma^{TPG}$ is set as 0.75 and $\beta^{TPG}$ set as 0.50 or higher. 

In terms of tidal delay, when $\Delta T$ is set as around $-3$ or $3$, higher daily average generation percentage can be achieved as shown in Fig. \ref{case2_curves_sufa}. 
The scheduling diagrams with $\Delta T$ set as $-3$, $0$ and $3$ and $\gamma^{TPG}$ set as $0$ and $0.50$ given a $\beta^{TPG}$ of $0.50$ can be found in Fig. \ref{case2_operation_figures}. 
It is obvious that when setting $\Delta T$ as $-3$, most of the generation from TPG lies in the middle of the day, aligning with peak of electricity load demand. 
While when setting $\Delta T$ as $+3$, generation of TPG shows two peaks, perfectly filling in time when generation from PV is not abundant, hence the complementarity helps to enhance consumption of RESs, especially TPG in this case. 

\subsection{Analysis of Uncertainty from Both Load Demand and Tidal Height} 
In this section, uncertainties of load demand and tidal height are considered concurrently, and the resulting investment decisions are analyzed and discussed. 

\begin{figure}[htbp]
    \centering
    \includegraphics[width=3.3in]{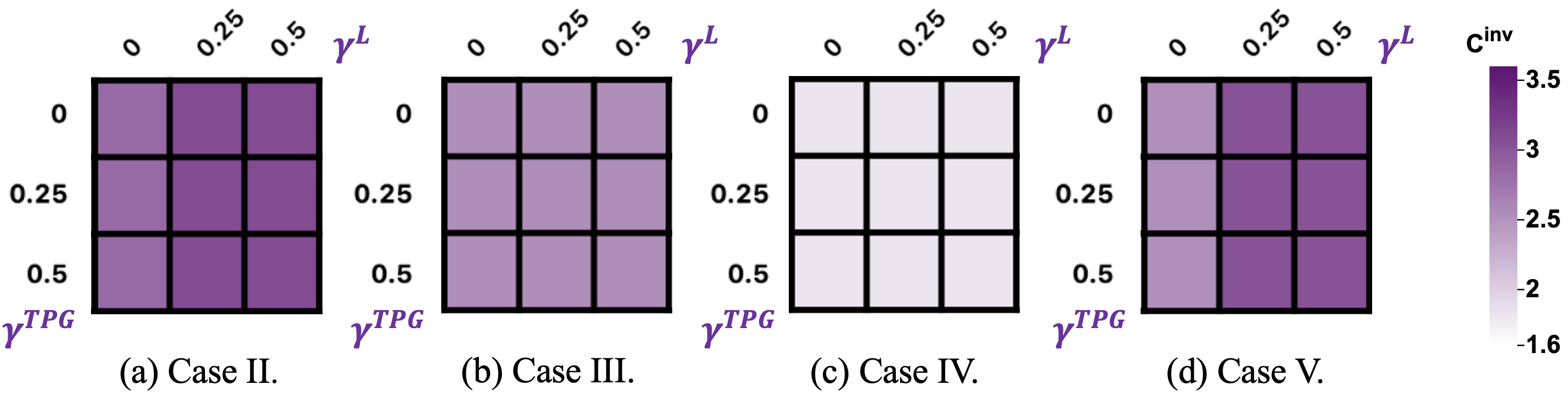} 
    \caption{Investment costs considering different $\gamma^{L}$ and $\gamma^{TPG}$ under different DER configurations. }
    \label{case3_heatmap_combo}
\end{figure}

The two uncertainty budget coefficients $\gamma^L$ and $\gamma^{TPG}$ are both taken from $0$ to $0.5$ with an interval of $0.25$, and the resulting operation costs are illustrated in Fig. \ref{case3_heatmap_combo}, where higher costs tend to make the color of the cell darker, and lower costs otherwise. 
Horizontal and vertical axes in the figure stand for uncertainty budget coefficient of load demand $\gamma^{L}$ and uncertainty budget coefficient of tidal level $\gamma^{TPG}$, respectively. 

As can be observed in Fig. \ref{case3_heatmap_combo}, both uncertainty budget coefficients give rise to $C^{inv}$, while $\gamma^{L}$ contributes more compared with $\gamma^{TPG}$. 
This is caused by the fact that $\gamma^{L}$ affects the load demand directly, while $\gamma^{TPG}$ is just a parameter that could affect which DER to generate the power, having less impacts on costs. 

Investment decisions of the cases when both $\gamma^L$ and  $\gamma^{TPG}$ are set as $0.5$ are listed in TABLE \ref{table_result_decision_case3}, where - indicates that the corresponding set of device is not allowed to be invested according to the setting of the case. 

\begin{table}[htp]
\renewcommand{\arraystretch}{1.1}
\caption{Investment Decisions and Costs under Different DER Configurations with $\gamma^L=\gamma^{TPG}=0.5$}
\label{table_result_decision_case3}
\centering
\begin{tabular}{c|cccc|c}
\hline
\multirow{2}{*}{Case \#} & \multicolumn{4}{c|}{Installed Capacity (MW)} & $C^{inv}$\\ \cline{2-5}
~ & DU & NDU & ESS & TPG & (M \$)\\ \hline
II & 21 & 6 & - & 14 & 3.098 \\
III & - & 6 & 6 & 14 & 2.544 \\
IV & - & 6 & - & 14 & 1.824 \\
V & 21 & - & 6 & 14 & 3.038 \\
VI & 21 & 6 & 3 & - & 2.294 \\
\hline
\end{tabular}
\end{table}

\section{Conclusion}
In this paper, a two-stage robust planning model for offshore microgrid incorporated with modeling of tidal power generation and seawater desalination units is proposed. 
The uncertainties of load demand and tidal power generation are both modeled and considered. 
The planning model is solved with the C\&CG algorithm. 
Case studies verify the model's effectiveness and analyze the robustness of different distributed energy resources configuration strategies. 
Besides, effects of uncertainties are also analyzed. 

The simulation results show the function of energy storage systems in peak shaving, helping to move the load peak and smoothen the load curve. 
Also, the microgrid can also work without traditional dispatchable units, hence a 100\% renewable microgrid could be possible. 
The outputs from tidal generators and PVs show complementarity due to the renewable energy sources' natural characteristics, and the power consumption of seawater desalination units tends to align with the outputs of renewable energy sources.  
Thus, the combination of them is a valuable part of offshore microgrid planning and worth spreading. 

However, the investment costs for tidal generation units are still high compared with traditional dispatchable units, and economic issues could lead to more installations of traditional dispatchable units that are less environmental-friendly. 
We are looking forward to seeing more developments in tidal power generation units, such as higher energy efficiency and more mature manufacturing, so that the investment costs could be driven down. 
Lower costs would lead to more investment of these environmental-friendly devices, which are beneficial for the environment while satisfying electricity load demand on the islands. 

The future work will focus on considering the connection between several microgrids and developing accelerating solving algorithm for the model. 

\ifCLASSOPTIONcaptionsoff
  \newpage
\fi

\bibliographystyle{IEEEtran}

\bibliography{mpp_revision_again_clean}

\end{document}